\documentclass{vldb}

\usepackage{balance}
\usepackage{listings}
\usepackage{color}
\usepackage{times}
\usepackage{url}
\usepackage{hyperref}

\definecolor{light-gray}{gray}{0.98}
\definecolor{dark-gray}{gray}{0.5}
\lstMakeShortInline^

\def\aroundlstskip{0ex}
\lstdefinestyle{noskip}{aboveskip=\aroundlstskip, belowskip=\aroundlstskip, abovecaptionskip=\aroundlstskip, belowcaptionskip=\aroundlstskip}
\lstnewenvironment{code}{}{}

\lstnewenvironment{code-segment}{\lstset{frame=TRBL}}{}

\usepackage{dashrule}

\newcounter{querycnt}
\lstnewenvironment{query}[2][]{
  
  \setcounter{lstlisting}{\value{querycnt}}
  \lstset{style=noskip, captionpos=b, float=h!t, caption=#1, label=#2, linewidth=3.45in}
}{
  \addtocounter{querycnt}{1}
}

\newcommand{\cde}[1]{\lstinline{#1}}

\usepackage{enumitem}  
\setlist[itemize]{style=multiline,noitemsep}

\newlength{\figurewidth}
\newlength{\columnfigurewidth}
\newenvironment{performancefigures}{
    \begin{figure*}[!ht]
    \vspace {0.18in}
    \centering
    \includegraphics[width=\figurewidth]
}{
    \end{figure*}
}

\begin{document}

\title{Apache VXQuery: A Scalable XQuery Implementation}

\numberofauthors{5}
\author{
\alignauthor
E. Preston Carman, Jr.\\
       \affaddr{UC Riverside}\\
       \email{ecarm002@ucr.edu}
\alignauthor 
Till Westmann\\
       \affaddr{Oracle Labs}\\
\alignauthor 
Vinayak R. Borkar\\
       \affaddr{UC Irvine}\\
\and 
\alignauthor 
Michael J. Carey\\
       \affaddr{UC Irvine}\\
\alignauthor
Vassilis J. Tsotras\\
       \affaddr{UC Riverside}\\
}

\maketitle
\begin{abstract}
The wide use of XML for document management and data exchange has created the need to query large repositories of XML data. 
To efficiently query such large data collections and take advantage of parallelism, we have implemented Apache VXQuery, an open-source scalable XQuery processor.
The system builds upon two other open-source frameworks -- Hyracks, a parallel execution engine, and Algebricks, a language agnostic compiler toolbox.
Apache VXQuery extends these two frameworks and provides an implementation of the XQuery specifics (data model, data-model dependent functions and optimizations, and a parser).
We describe the architecture of Apache VXQuery, its integration with Hyracks and Algebricks, and the XQuery optimization rules applied to the query plan to improve path expression efficiency and to enable query parallelism.
An experimental evaluation using a real 500GB dataset with various selection, aggregation and join XML queries shows that Apache VXQuery performs well both in terms of scale-up and speed-up.  
Our experiments show that it is about 3x faster than Saxon (an open-source and commercial XQuery processor) on a 4-core, single node implementation, and around 2.5x faster than Apache MRQL (a MapReduce-based parallel query processor) on an eight (4-core) node cluster.
\end{abstract}

\section{Introduction}\label{sec:intro}
The widespread acceptance of XML as a standard for document management and data exchange has enabled the creation of large repositories of XML data. 
To efficiently query such large data collections, a scalable implementation of XQuery is needed that can take advantage of parallelism. 
While there are various native open-source XQuery processors (Saxon \cite{kay2004saxon}, Galax \cite{fernandez2003implementing}, etc.) they have been optimized for single-node processing and do not support scaling to many nodes. 
To create a scalable XQuery processor, one could: 1) add scalability to an existing XQuery processor, 2) start from scratch, or 3) extend an existing scalable query framework to support XQuery. 
Unfortunately, existing XQuery processors would require extensive rewriting of their core architecture features to add parallelism. 
Similarly, building an XQuery processor from scratch would involve the same complex scalable programming (some unrelated to the XML data model). 
The last option, extending an existing scalable framework to support XQuery, seems advantageous since it combines the benefits of proven parallel technology with a shorter time to implementation.

Among the several scalable frameworks available, one could use a relational parallel database engine and take advantage of its mature optimization techniques. 
However, this entails the overhead of translating the data/queries to the relational model and back to XML; moreover, long XML path queries may result in many joins. 
Another approach is to build an XQuery processor on top of the MapReduce \cite{Dean:2004:MSD:1251254.1251264} framework. 
Examples include ChuQL \cite{khatchadourian2011having}, which is a MapReduce extension to XQuery built on top of Hadoop \cite{borthakur2007hadoop}, and HadoopXML \cite{choi2012hadoopxml}, which combines many XPath queries into a few Hadoop MapReduce jobs. 
Similarly, Apache MRQL \cite{fegaras2011xml} translates XPath queries into an SQL-like language implemented through MapReduce operators. 
However, these Hadoop-based approaches are limited in that they can only use the few MapReduce operators available (i.e. map, reduce, and combine).
 
Recently, frameworks have been proposed that generalize the MapReduce execution model by supporting a larger set of operators to create parallel jobs (including Hyracks \cite{borkar2011hyracks}, Spark \cite{zaharia2010spark}, and Stratosphere \cite{alexandrov2014stratosphere}). 
Such 'dataflow' systems \cite{babu2013massively} typically include flexible data models supporting a wide range of data formats (relational, semi-structured, text, JSON, XML, etc.) which makes them easy to extend. 
In this paper, we utilize Hyracks as our parallel framework and use Algebricks, a language agnostic compiler toolbox, to implement XQuery. 
 
Our implementation is available as open source at the Apache Software Foundation \cite{vxquery2014}. 
We have performed an experimental evaluation using a large (500GB) real dataset (a NOAA weather dataset from \cite{NOAA-GHCND}) and various selection, aggregation and join XML queries that show the efficiency of our XQuery processor, both in terms of speed up and scale up.

The rest of this paper is organized as follows: Section \ref{sec:work} reviews current approaches for querying large XML data repositories while Section \ref{sec:stack} covers the VXQuery software stack with details about the underlying framework (Hyracks and Algebricks) and how the data model, parser, and runtime were extended for XQuery support. 
Given the specifics of XQuery, we had to extend existing Algebricks rewrite rules and introduce new ones; this discussion appears in Section \ref{sec:rules}.
Finally, Section \ref{sec:performance} presents the results of our experiments on VXQuery's performance as well as a comparison with two existing open-source XML processors -- a single threaded one, SaxonHE, and a parallel one, Apache MRQL.

\newpage
\section{Related Work}\label{sec:work}
Hadoop \cite{hadoop2011apache} provides a framework for distributed processing based on the MapReduce model. That leaves a significant implementation burden on the application programmer. As a result, a number of languages have been proposed on top of Hadoop (e.g. Hive \cite {thusoo2009hive}, PigLatin \cite{olston2008pig}, and Jaql \cite{beyer2011jaql}); however, popular high-level MapReduce languages do not support the XML data model.
Recent approaches to close this gap include: ChuQL, Apache MRQL, HadoopXML, and Oracle XQuery for Hadoop.

ChuQL \cite{khatchadourian2011having} extends XQuery to include MapReduce support for processing native XML on Hadoop. 
In ChuQL, a MapReduce expression is included as an XQuery function, allowing the query writer to specify the MapReduce job definition in XQuery. 
In contrast, VXQuery hides all parallel processing details from the query writer while still using standard XQuery constructs.

Apache MRQL (MapReduce Query Language) \cite{fegaras2012mrql, fegaras2011xml} is a SQL-like language designed to run big data analysis tasks. 
The language supports parsing XML data from Hadoop through a \textit{source} expression that defines the XML parser, XML file, and XML tags.
The XML parser processes the XML file and returns all the elements matching these tags;
Apache MRQL then translates these elements into the Apache MRQL data model.
Each query is translated to an algebra expression for the Apache MRQL cost-based optimizer, which builds upon known relational query and MapReduce optimization techniques.
The algebra uses a small number of physical operators to create a more efficient MapReduce job than directly writing it using the MapReduce operators. 

HadoopXML \cite{choi2012hadoopxml} processes a single large XML file with a predetermined set of queries (each currently in a subset of XPath). 
The engine identifies the query commonalities (paths that are common) and executes those once; it then shares the common results and augments them with the non-common parts per query.
This processing is performed using MapReduce jobs.
When a query is executed, the query optimizer determines the optimum number of jobs to execute the requested query. 

Recently, Oracle released Oracle XQuery for Hadoop (OXH) \cite{oracle2014hadoop}, which runs XQuery data transformations by translating them into a series of MapReduce jobs. 

In summary, the above approaches share the MapReduce framework and are thus limited to using only the available MapReduce operators.
VXQuery differs in that it is built on top of a more general scalable framework (Hyracks) and can match XQuery computational tasks to Hyracks' richer existing operators; this in turn provides better performance. 
As will be seen in our experimental section, our rewrite rules, together with Hyracks' efficiency, provides over twice the performance of other approaches that perform XML processing on top of MapReduce.
Also related to our approach is PAXQuery \cite{camacho2014paxquery}, which implements XQuery on top of Stratosphere \cite{alexandrov2014stratosphere}, which is another dataflow system (like Hyracks).
PAXQuery was not publicly available as of the writing of this paper.
Similarly, the Apache MRQL group is currently working on supporting Apache MRQL on top of Apache Flink \cite{flink2014} (which evolved from the Stratosphere project) but at the time of this writing that implementation was still under development.

\section{Apache VXQuery's Stack}\label{sec:stack}
\begin{figure}
\center
\includegraphics[width=.65\columnwidth]{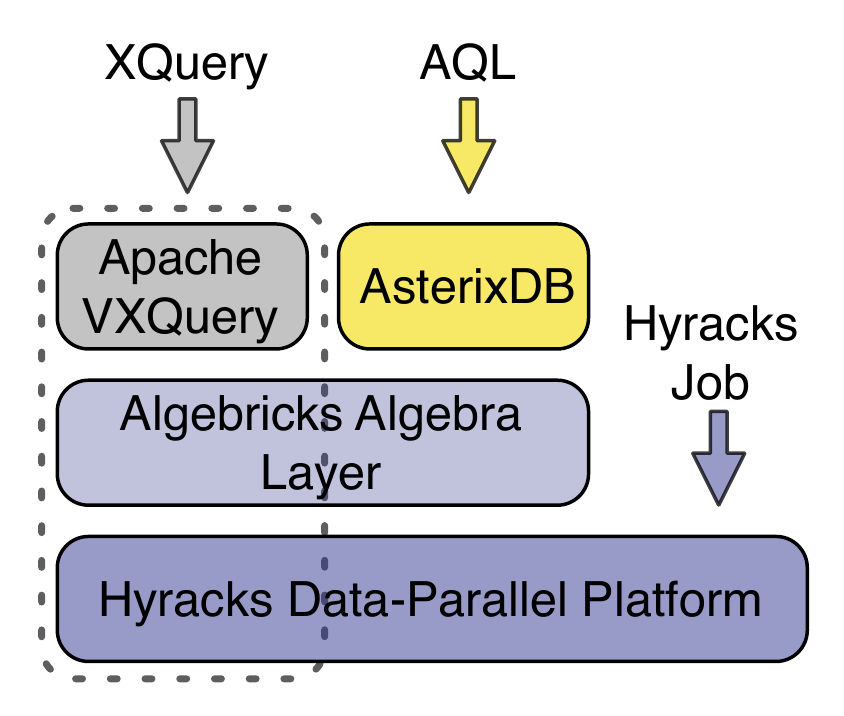}
\vspace{-1em}
\caption{The layers of the Apache VXQuery stack.}
\vspace{-1em}
\label{fig:vxquery_stack}
\end{figure}

Apache VXQuery's software stack can be represented in three layers, as shown in Figure \ref{fig:vxquery_stack}.
The top layer, Apache VXQuery, forms an Algebricks logical plan based on parsing a supplied XQuery.
The initial Algebricks logical plan is then optimized and translated into an Algebricks physical plan that maps directly to a Hyracks job.
The Hyracks platform executes the job and returns the results.
A brief explanation of each layer in the stack follows in the next subsection.
Figure \ref{fig:vxquery_stack} also shows AsterixDB \cite{alsubaiee2014asterixdb}, another system that uses the Hyracks and Algebricks infrastructure, 

\subsection{Hyracks}

Hyracks is a data-parallel execution platform that builds upon mature parallel database techniques and modern big data trends \cite{hyracks2014url, borkar2011hyracks}.
This generic platform offers a framework to run dataflows in parallel on a shared-nothing cluster.
The system was designed to be independent of any particular data model.
Hyracks processes data in partitions of contiguous bytes, moving data in fixed-sized frames that contain physical records, and it defines interfaces that allow users of the platform to specify the data-type details for comparing, hashing, serializing and de-serializing data.
Hyracks provides built-in base data types to support storing data on local partitions or when building higher level data types (first row of Table \ref{tab:data_types}).

A Hyracks job is defined by a dataflow DAG with operators (nodes) and connectors (edges).
During execution, the operators allow the computation to consume an input partition and produce an output partition while the connectors redistribute data among partitions.
The dataflow among Hyracks operators is push-based -- each source (producer) operator pushes the output frames to a target (consumer) operator.
The extensible runtime platform provides a number of operators and connectors for use in forming Hyracks jobs.
While each operator's operation is defined by Hyracks, the operator relies on data-model specific functionality provided by the user of the platform.


\subsection{Algebricks}

Algebricks \cite{algebricks2014, borkar2013algebricks} is a parallel framework providing an abstract algebra for parallel query translation and optimization. 
This language-agnostic toolbox complements the lower-level extensible Hyracks platform.
Implementations of data-intensive languages can extend its model-agnostic algebraic layer to create parallel query processors on top of the Hyracks platform.
A language developer is free to define the language and data model when using the Hyracks platform and the Algebricks toolkit.
Algebricks features a rule-based optimizer and data model neutral operators that each allow for language specific customization.

A system that uses Algebricks for its query processing begins by lexically-analyzing and parsing the query.
The parsed query is then translated into a dataflow DAG, or query plan, using Algebricks' logical operators as an intermediate language or algebra.
The Algebricks rule-based optimizer consists of rewrite rules that transform the plan into an improved and equivalent plan.
The optimizer modifies the query plan over three stages.
The first is a Logical-to-Logical plan optimizer that creates alternate logical plan formations.
Once the logical plan is finalized, the Logical-to-Physical plan optimizer converts the logical operators into a physical plan.
Then, the physical optimizer considers the operator characteristics, partition properties, and data locality to choose the optimal physical implementation for the plan.
Algebricks provides generic language-independent rewrite rules for each stage and allows for the addition of other rules. 
Finally, a Hyracks job is generated and submitted for execution on a Hyracks cluster.

Algebricks' intermediate logical algebra uses logical operators that map onto Hyracks' physical operators.
A logical operator's properties are considered when determining the best physical operator.
For example, a join query that has an equijoin predicate allows a hash based join instead of the default nested loop join.
The Algebricks logical operators exchange data in the form of logical tuples, each of which is a set of fields.
The field names are represented by \$\$ followed by a number in remaining text.
The following Algebricks logical operators are commonly used in VXQuery: 

\begin{itemize}
  \item The DATASCAN operator reads from a data source and returns one tuple for each item in the data source. 
  \item The ASSIGN operator executes a scalar expression on a tuple and adds the result as a new field in the tuple.
  \item The DISTRIBUTE-RESULT operator collects the designated variable for the query result, once the job is completed the controller will request it and return it to the user.
  \item The EMPTY-TUPLE-SOURCE operator initializes the first tuple without any fields. 
Algebricks uses this operator to start all DAG dataflow paths.
  \item The JOIN operator matches and combines tuples from two streams of input tuples.
  \item The AGGREGATE operator executes an aggregate expression to create a single result tuple from a stream of input tuples. 
The result is held until all tuples are processed and then returned in a single tuple.
  \item The UNNEST operator executes an unnesting expression for each tuple, creating a stream of single item tuples from each. 
  \item The SUBPLAN operator executes a nested plan for each tuple input.
  \item The NESTED-TUPLE-SOURCE operator is used as the leaf operator in nested plans.
The operator links the nested plans with the input to the operator (such as a SUBPLAN) defining the nested plan.
\end{itemize}

The Algebricks operators are each parameterized with custom expressions.
The expressions map directly to specific language functions or support runtime features.
Each expression is an instance of one the following expression types: scalar, aggregate, and unnesting.
Most operators use a scalar expression, while the AGGREGATE and UNNEST operators have their own expression types.
The three expression types differ in their input and output cardinalities.
Scalar expressions operate on a single value and return a single value.
Aggregate expressions consume many values to create a single result. 
Unnesting expressions consume a single (usually structured) value to create many new values.
Correspondingly, the AGGREGATE and UNNEST operators change the cardinality of the tuple stream.

\subsection{Apache VXQuery}
\label{sec:apache_vxquery}

Apache VXQuery extends the language agnostic layer provided by Algebricks to create a scalable XQuery processor.
VXQuery provides a binary representation of the XQuery Data Model (XDM), an XQuery parser, an XQuery optimizer, and the data model dependent expressions.
The XML data supplied to XQuery comes in the form of non-fragmented XML documents partitioned evenly throughout a cluster. 
A SAX based XML parser translates the XML documents at runtime into XDM instances.
Hyracks base types were extended to build untyped XDM instances for the XQuery node types and the XQuery atomic types. 
(All XQuery types used are listed in Table \ref{tab:data_types}.) 

\begin{table}
\vspace {0.2in}
\centering
\renewcommand{\arraystretch}{1.2}
\begin{tabular}{|l|l|l|} \hline
Hyracks Base & boolean, byte, short, integer, long, \\
 & double, float, UTF8 string \\ \hline
XQuery Node & attribute, comment, document, element, \\
 & processing instruction, text \\ \hline
XQuery Atomic & binary, decimal, date, datetime, time, \\
 & duration, QName \\ \hline
\end{tabular}
\caption{Apache VXQuery builds on the Hyracks Base types to create the XQuery Node and Atomic data types.}
\label{tab:data_types}
\end{table}

Query evaluation proceeds through the usual steps.
The query is parsed into an abstract syntax tree (AST) and is then analyzed, normalized, and translated into a logical plan.
The logical plan consists of Algebricks data model independent operators parameterized with VXQuery data model dependent expressions.
The logical plan is then optimized using both generic rewrite rules provided by Algebricks and XQuery specific rewrite rules provided by VXQuery (discussed in Section \ref{sec:rules}).
After rewriting the logical plan, it is translated into a physical plan and optimized further (physical optimization includes, e.g., the selection of join methods or the distribution of the plan).
Finally the physical plan is translated into a Hyracks job that is executed.
Similar to Algebricks operators that have physical representations based on Hyracks operators, VXQuery provides executable functions that implement VXQuery's data model dependent expressions.

Special attention is required regarding how the XQuery data model (XDM) defines a set of items as a sequence.
In VXQuery, an XDM sequence can have two forms: a \emph{sequence item} or a \emph{tuple stream}.
A sequence item  holds all the values in a single tuple field;
a tuple stream represents the sequence using a field with the same name in multiple tuples.
To provide a framework for switching between these representations, we implemented two internal runtime expressions, namely, the \textit{iterate} unnesting expression and the \textit{create\_sequence} aggregate expression. 
The \textit{iterate} unnesting expression works with Algebricks' UNNEST operator to convert a tuple field that holds an sequence item into a stream of individual tuples. 
The \textit{create\_sequence} aggregate expression executes within Algebricks' AGGREGATE operator to consume a tuple stream and consolidate it into a sequence item for inclusion in a single output tuple. 
The two expressions are used during the logical rewrite process to switch between formats to enable further optimization rules to be applied to the query plan.

\begin{figure}
\includegraphics[width=\columnwidth]{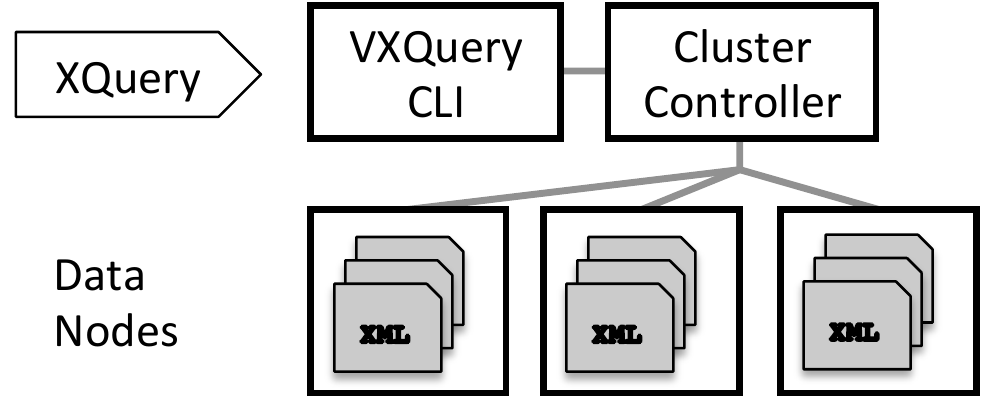}
\vspace{-1em}
\caption{The VXQuery cluster configuration.}
\vspace{-1em}
\label{fig:vxquery_cluster}
\end{figure}

At runtime, the VXQuery cluster processes a query using the VXQuery Client Library Interface (CLI), a Hyracks Cluster Controller, and some Hyracks Data Nodes (as shown in Figure \ref{fig:vxquery_cluster}). 
The process starts with a user submitting an XQuery statement to the VXQuery CLI for parallel execution.
The CLI parses and optimizes the query and submits the generated Hyracks job to the cluster controller, which manages and distributes tasks to each of the data nodes for evaluation.
Each data node contains XML files, an XML parser and the XQuery runtime expressions used to evaluate the node's tasks.
Finally, the cluster controller collects the data nodes' results and sends the result back to the VXQuery CLI, which returns the result to the user.


\section{Rewrite Rules}\label{sec:rules}

Algebricks provides generic rules for both Logical-to-Logical and Logical-to-Physical plan optimizations.
These rules include actions that consolidate, push down, and/or remove operators based on the operators' properties and the query plan. 
In addition, to build the XQuery optimizer we needed to implement XQuery-specific rules; these rules fall into two categories.
The {\em Path Expression Rewrite Rules} attempt to remove subplans that are introduced by the unnesting required to evaluate path expressions. 
The {\em Parallel Rewrite Rules} transform the plan to enable parallel evaluation for specific XQuery constructs (aggregation, join, and data access) using both pipelined and partitioned parallelism.

\subsection{Path Expression Rewrite Rules}

The normalization phase of query translation introduces explicit operations into the query plan that ensure the correctness of the plan (for example sorting to maintain document order).
However, some of these operations may not be required based on knowledge of the structure of the plan and the implementation of operators and expressions.
The following XML segment is based on the sample XML tree from the W3Schools tutorial \cite{w3schools2014xml} for XQuery and will be used to outline the path expression rewrite rules.

\lstset{language=XML,keepspaces=true,breakatwhitespace=false,breaklines=true}
\begin{lstlisting}
<?xml version="1.0" encoding="ISO-8859-1"?>
<bookstore>
  <book id="1" category="COOKING">
    <title lang="en">Everyday Italian</title>
    <author>Giada De Laurentiis</author>
    <year>2005</year>
  </book>
  <book id="2" category="CHILDREN">
    <title lang="en">Harry Potter</title>
    <author>J K. Rowling</author>
    <year>2005</year>
  </book>
  ...
</bookstore>
\end{lstlisting}

Consider the following simple query. 

\begin{lstlisting}
doc("book.xml")/bookstore/book
\end{lstlisting}

The query reads data from the document \textit{book.xml} located
in the file system using the XQuery \textit{doc} function. 
Next, the first child path step expression ("/bookstore") is applied to the document node.
Three stages are used when applying the child path step expression to a tuple: each input node is iterated over, any matching child nodes are put into a single sequence, and the sequence is then sorted in document order. 
The same path step process is applied to each resulting "bookstore" element node for the second child path step expression ("/book").
Finally, each "book" element node is then returned in the final query result. 

\begin{figure*}
\centering
\epsfig{file=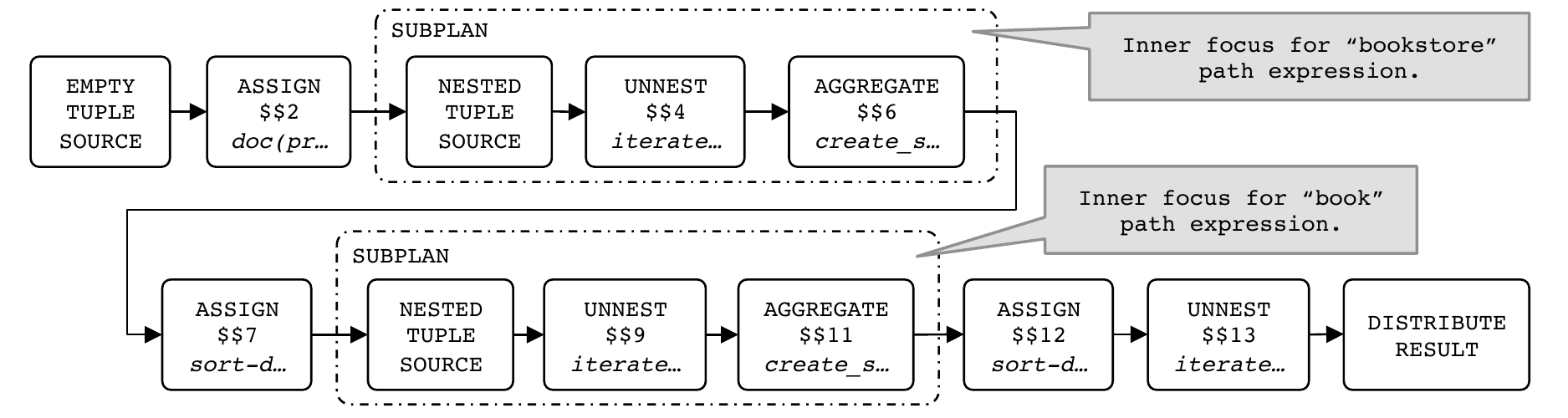, width={.9\textwidth}}
\vspace{-1em}
\caption{Example query plan before applying rewrite rules.}
\label{fig:query_plan_example}
\end{figure*}

The VXQuery query processor creates the initial plan shown below (after removing unused variables); here the curly braces represent nested plans that are executed for each of the SUBPLAN's input tuples. 
Schematically this plan corresponds to the DAG in Figure \ref{fig:query_plan_example}. 
This DAG is a single path of execution and can be represented by a list of operators in reverse dataflow order.
By default, each DAG is initialized with an EMPTY-TUPLE-SOURCE operator and collects its results into a DISTRIBUTE-RESULT operator.

\begin{lstlisting}
DISTRIBUTE-RESULT( $$13 )
UNNEST( $$13:iterate($$12) )
ASSIGN( $$12:sort-distinct-nodes-asc-or-atomics($$11) )
SUBPLAN {
  AGGREGATE( $$11:create_sequence(child(treat($$9, element_node), "book")) )
  UNNEST( $$9:iterate($$7) )
  NESTED-TUPLE-SOURCE
}
ASSIGN( $$7:sort-distinct-nodes-asc-or-atomics($$6) )
SUBPLAN {
  AGGREGATE( $$6:create_sequence(child(treat($$4, element_node), "bookstore")) )
  UNNEST( $$4:iterate($$2) )
  NESTED-TUPLE-SOURCE
}
ASSIGN( $$2:doc(promote(data("books.xml"), string) )
EMPTY-TUPLE-SOURCE
\end{lstlisting}

\begin{figure*}[t]
\centering
\begin{tabular}{ccc}
\includegraphics[width=\columnfigurewidth]{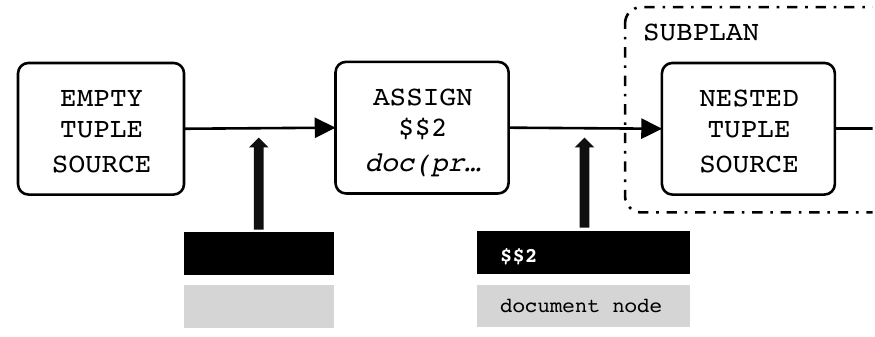}
&&
\includegraphics[width=\columnfigurewidth]{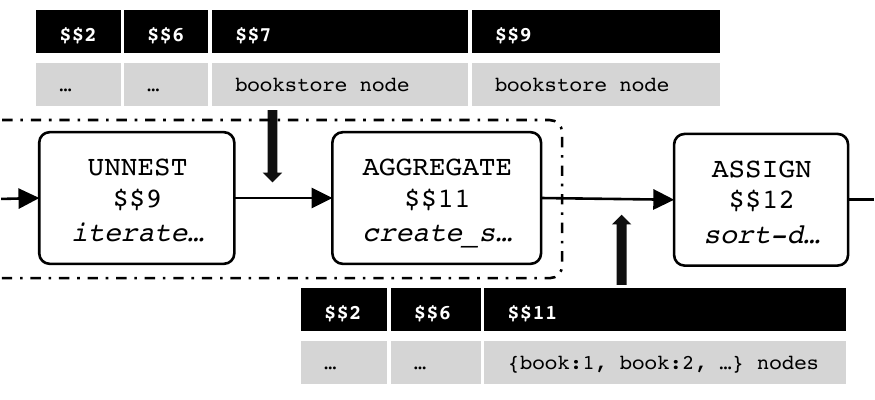}
\\
\textbf{(a) First ASSIGN operator.}
&&
\textbf{(b) Last AGGREGATE operator.}
\\[2ex]
\includegraphics[width=\columnfigurewidth]{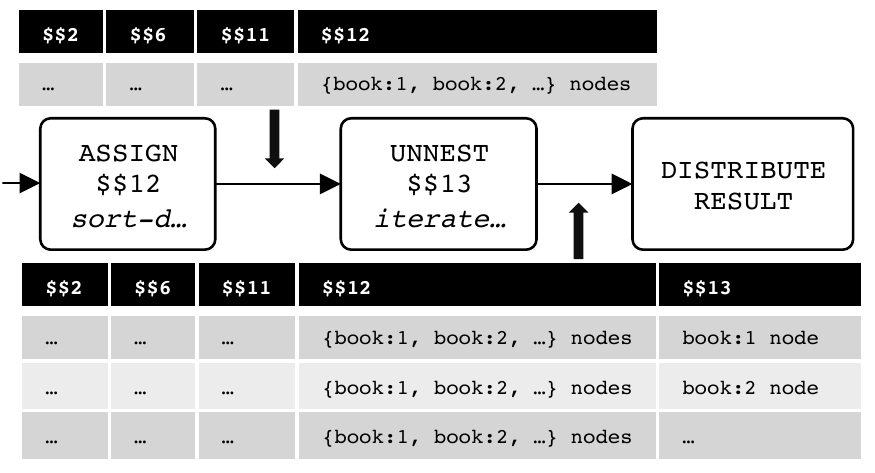}
&&
\includegraphics[width=\columnfigurewidth]{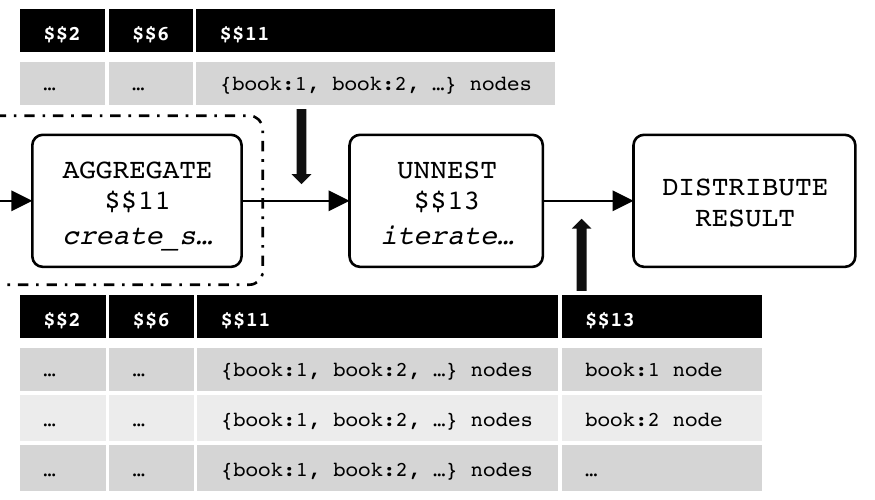}
\\
\textbf{(c) Last UNNEST operator.}
&&
\textbf{(d) Remove SUBPLAN Operators rule.}
\\
\end{tabular}
\caption{Dataflow segments}
\vspace{-1em}
\label{fig:data_flow_segments}
\end{figure*}

The EMPTY-TUPLE-SOURCE operator creates the initial empty tuple. 
The \textit{doc} expression in the ASSIGN \$\$2 operator returns a document node using the string URI argument and adds a new field -- \$\$2:document node -- to the tuple, as shown in Figure \ref{fig:data_flow_segments}(a).
The \textit{promote} and \textit{data} expressions ensure the \textit{doc} URI argument will be a string. 
The SUBPLAN operator above the ASSIGN \$\$2 operator uses a nested plan to implement the first and second stages of the \cde{/bookstore} path step. 
The subplan's nested plan ensures the correct dynamic context for the path step and provides an "inner focus" to evaluate the expression on each item in the sequence for the next step (if any).
The NESTED-TUPLE-SOURCE operator connects the nested plan to the SUBPLAN's input dataflow. 
The input tuple is passed on to the UNNEST \$\$4 operator where each \$\$2:document item is iterated over and added as \$\$4:document.
For the root path step expression, there is only one item in the sequence.  
The inner focus is closed with AGGREGATE \$\$6 processing all SUBPLAN tuples using the \cde{create\_sequence(child(treat(\$\$4, element\_node), "bookstore"))} expression.
The expression ensures that \textit{child} expression's argument is of type "element\_node" (\textit{treat}), finds all "bookstore" child nodes (\textit{child}), and creates a sequence of all the results (\textit{create\_sequence}). 
The resulting tuple now holds two fields: \$\$2:document node and \$\$6:"bookstore" node. 
All the SUBPLAN variables are discarded except the final operator's result (in this case, AGGREGATE \$\$6).
The third stage of the path step is completed though the ASSIGN \$\$7 with \cde{sort-distinct-nodes-asc-or-atomics(\$\$6)}.
The expression creates a new field with nodes that are in document order and duplicate free from \$\$6:"bookstore" node.
Since there is only one item, the "bookstore" node is copied over to \$\$7.

The next SUBPLAN creates the inner focus for the \cde{/book} path step expression.
Similar to the \cde{/bookstore} path step, the nested plan iterates over the input tuples and saves all child nodes \{book:1, book:2, ...\} to \$\$11, as shown in Figure \ref{fig:data_flow_segments}(b).
The ASSIGN \$\$12 ensures document order in the child "book" node sequence by removing duplicates and sorting the sequence. 
Finally, each item in \$\$12:\{book:1, book:2, ...\} is unnested by UNNEST \$\$13 to create a tuple stream for the DISTRIBUTE-RESULT operator. 
See Figure \ref{fig:data_flow_segments}(c) for a graphic representation of the tuples before and after this UNNEST operator.

The initial query plan is inefficient and can be improved in several ways: we can (i) remove the computationally expensive sort operators (as document order is not changed by any of the other operators) and (ii) remove the SUBPLAN operators (since each SUBPLAN corresponds to a simple step expression, the inner focus is not required). 
After these optimization rules, the plan can be cleaned further by (iii) enabling unnesting  (improves operator efficiency) and (iv) merging the path step unnesting operators (reduces number of operators).
This is achieved by the rules described next.

\subsubsection{Remove Sort Expressions}

A path expression requires all results to be in document order and duplicate free. 
In a single XML file, the order is determined by the serialization of the XML document. 
When querying multiple documents, the result order must be stable during the evaluation of a query.
These properties are ensured by using the ASSIGN operator with the \textit{sort-distinct-nodes-asc-or-atomics} expression.
In VXQuery, properties of the data are tracked through the query plan using knowledge of each operator and expression (similar to \cite{fernandez2004automata}).
If the analysis reveals that the document order and/or duplicate free properties are intact at this point in the query plan, then the sort expression can be removed or replace.
The updated plan may include any of the following expressions based on the intact properties: \textit{sort-distinct-nodes-asc-or-atomics} (both sorting and duplicate elimination), \textit{sort-nodes-asc-or-atomics} (only sorting), \textit{distinct-nodes-or-atomics} (only duplicate elimination), or no expression.
For example, if the plan does not include operators or expressions that change the document order or introduce duplicates, the original ASSIGN with the sort operator is replaced by no-op operator and is subsequently removed. 
As a result, the updated example query plan becomes:

\begin{lstlisting}
DISTRIBUTE-RESULT( $$13 )
UNNEST( $$13:iterate($$11) )
SUBPLAN {
  AGGREGATE( $$11:create_sequence(child(treat($$9, element_node), "book")) )
  UNNEST( $$9:iterate($$6) )
  NESTED-TUPLE-SOURCE
}
SUBPLAN {
  AGGREGATE( $$6:create_sequence(child(treat($$4, element_node), "bookstore")) )
  UNNEST( $$4:iterate($$2) )
  NESTED-TUPLE-SOURCE
}
ASSIGN( $$2:doc(promote(data("books.xml"), string) )
EMPTY-TUPLE-SOURCE
\end{lstlisting}

\subsubsection{Remove SUBPLAN Operators}
\label{SBR}


The path expressions in our example do not require the inner focus created by the SUBPLAN operator (since we simply find all the children book nodes of a bookstore node, without any further qualification per path step).
Having eliminated the sort expression and without the inner focus, the two internal VXQuery expressions
for switching tuple representations, \textit{create\_sequence} and \textit{iterate},
are now neighbors, so the tuple translation is unnecessary.
That is, the sequence built by \cde{create\_sequence(child(treat(\$\$9, element\_node), "book"))} is immediately broken up into individual items in the \textit{iterate} expression.
Figure \ref{fig:data_flow_segments}(d) graphically shows the tuple stream before and after the \textit{iterate} expression.
The \cde{iterate(\$\$11)} expression has created a new tuple and field for each item in the \$\$11 sequence and labeled the new field \$\$13.
We could avoid creating the materialized sequence by removing the AGGREGATE operator and directly feed individual tuples as they come.
Once the AGGREGATE operator is removed, the SUBPLAN that serves as the container for the aggregate is not needed either.
Updating our example with the application of the Remove SUBPLAN Operators rule results in:

\begin{lstlisting}
DISTRIBUTE-RESULT( $$13 )
UNNEST( $$13:iterate($$11) )
ASSIGN( $$11:child(treat($$9, element_node), "book") )
UNNEST( $$9:iterate($$6) )
SUBPLAN {
  AGGREGATE( $$6:create_sequence(child(treat($$4, element_node), "bookstore")) )
  UNNEST( $$4:iterate($$2) )
  NESTED-TUPLE-SOURCE
}
ASSIGN( $$2:doc(promote(data("books.xml"), string) )
EMPTY-TUPLE-SOURCE
\end{lstlisting}

The updated plan again shows a \textit{create\_sequence} consumed by a \textit{iterate} expression, thus the rule can be applied a second time:

\begin{lstlisting}
DISTRIBUTE-RESULT( $$13 )
UNNEST( $$13:iterate($$11) )
ASSIGN( $$11:child(treat($$9, element_node), "book") )
UNNEST( $$9:iterate($$6) )
ASSIGN( $$6:child(treat($$4, element_node),"bookstore") )
UNNEST( $$4:iterate($$2) )
ASSIGN( $$2:doc(promote(data("books.xml"), string) )
EMPTY-TUPLE-SOURCE
\end{lstlisting}

A byproduct of removing the \textit{create\_sequence} expression is that the plan will now use smaller tuples between operators. 
In general, to apply this rule we are searching for the following query segment within a plan (query segments are shown with a double border):

\begin{code-segment}
UNNEST( $result:iterate($sequence) )
SUBPLAN{
  AGGREGATE( $sequence:create_sequence(@exp0) )
  @NESTED
  NESTED-TUPLE-SOURCE
}
\end{code-segment}

This rule replaces the combination of the SUBPLAN, AGGREGATE, and \textit{create\_sequence} with the \cde{@exp0} expression as the input to an ASSIGN operator. 
Once the AGGREGATE and the SUBPLAN operators have been removed, the nested plan (denoted as \cde{@NESTED}) is merged into the remaining query plan.
The \textit{iterate} expression is kept to split up any sequences produced by the \cde{@exp0} expression.
The old query segment is replaced by the equivalent one seen below:

\begin{code-segment}
UNNEST( $result:iterate($exp_result) )
ASSIGN( $exp_result:@exp0 )
@NESTED
\end{code-segment}

\subsubsection{Replace Scalar with Unnesting Expressions}

The Remove Sort Expressions and Remove SUBPLAN Operator rules revealed clean up opportunities.
Each child step expression is now within an ASSIGN operator, which is a scalar operator. 
As a result, the book child step expression (\$\$11) will read a bookstore node and return all its book child nodes in a single tuple.
This is then fed to the UNNEST operator (\$\$13) whose iterate expression will return each book child node in a separate tuple.
Instead, we can merge the ASSIGN \textit{child} (\$\$11) and UNNEST \textit{iterate} (\$\$13) operators into an UNNEST \textit{child} operator and return each book child node, as it is found, in a separate tuple.
Note that the \textit{child} expression for the UNNEST operator is now an unnesting implementation of the expression.
Below we depict the example query plan after applying this rewrite rule for both the book and bookstore path expressions:

\begin{lstlisting}
DISTRIBUTE-RESULT( $$13 )
UNNEST( $$13:child(treat($$9, element_node), "book") )
UNNEST( $$9:child(treat($$4, element_node),"bookstore") )
UNNEST( $$4:iterate($$2) )
ASSIGN( $$2:doc(promote(data("books.xml"), string) )
EMPTY-TUPLE-SOURCE
\end{lstlisting}

A byproduct of this rule is that the new query plan opens up opportunities for pipelining since each item is passed on as soon as the result is ready. 
This approach can be generalized to find any scalar expression that has an equivalent unnesting expression.
To apply this rule, the generic query segment that we search for within a plan is as follows:

\begin{code-segment}
UNNEST( $result:iterate($scalar_result) )
ASSIGN( $scalar_result:scalar_expression(@expression) )
\end{code-segment}

As the rewrite rule removes the ASSIGN operator, \$scalar\_result must not be used elsewhere in the plan. 
The replacement query segment is:

\begin{code-segment}
UNNEST( $result:unnesting_expression(@expression) )
\end{code-segment}

\subsubsection{Combine Path Expression UNNEST operators}

After applying these path expression rewrite rules, the sample query plan only uses only one UNNEST operator to represent each path expression.
The input for the book path expression is the output of bookstore path expression.
Instead of doing each path expression in separate operators, the two expressions can be merged into a single UNNEST.
The variable \$\$9 is replaced by the bookstore expression and the bookstore UNNEST operator is removed.
After applying this rewrite rule, the sample query plan is:

\begin{lstlisting}
DISTRIBUTE-RESULT( $$13 )
UNNEST( $$13:child(treat(child(treat($$4, element_node), "bookstore"), element_node), "book") )
UNNEST( $$4:iterate($$2) )
ASSIGN( $$2:doc(promote(data("books.xml"), string) )
EMPTY-TUPLE-SOURCE
\end{lstlisting}

As seen in our examples, the path expression rewrite rules can be applied multiple times to optimize a series of path expressions.

\subsection{Parallel Rewrite Rules}

After the path expression rewrite rules are applied, the plan is further optimized for parallel XQuery processing. 
Hyracks allows for both pipelined and partitioned parallelism.
We hence introduce rules to enable the use of Hyracks' parallel execution features.
To take advantage of pipelining in VXQuery we create fine grained data items; for example, the next rule should not compute a whole collection at once, but should instead compute chunks that can be fed to the remaining operators.
As a side effect, the needed buffer size (Hyracks' frame) is reduced between the operators in the pipeline. 
To introduce partitioned parallelism we use partitioned data access for physically partitioned data and we use partitioned parallel algorithms for join and aggregation.

\subsubsection{Introduce the DATASCAN Operator}

To query a collection of XML documents, XQuery defines a function called \textit{collection} that maps a string to a sequence of nodes.
VXQuery interprets the string as a directory location, reads in data from the files in the directory, and returns all nodes as a single sequence value.
Since the collection query considers many documents, it can produce a large number of query results.
Instead of gathering all nodes into a single sequence, we would like to send one node at a time through the pipeline. 
In addition, materializing such a large result sequence at runtime may exceed the Hyracks' frame size (which is configurable) causing the query to fail.
To avoid this problem, we combine the \textit{collection} expression with an \textit{iterate} expression (typically inserted because of a path step or a for clause) to split the large document sequence into many single document tuples, thus reducing the size of the materialized result.
Below is a sample \textit{collection} query similar to the previous single document query example, followed by the query plan generated after the path expression rules have been applied.

\begin{lstlisting}
collection("/books")/bookstore/book
\end{lstlisting}
	
\begin{lstlisting}
DISTRIBUTE-RESULT( $$13 )
UNNEST( $$13:child(treat(child(treat($$4, element_node), "bookstore"), element_node), "book") )
UNNEST( $$4:iterate($$2) )
ASSIGN( $$2:collection(promote(data("/books"), string) )
EMPTY-TUPLE-SOURCE
\end{lstlisting}

The path expression rules have conveniently moved an UNNEST \textit{iterate} above the ASSIGN \textit{collection}, creating a stream of XML document tuples. 
Algebricks offers a DATASCAN operator to directly create a stream of tuples based on a data source.
Since \textit{collection} already defines the data source, the DATASCAN operator can be used to replace UNNEST \textit{iterate} and ASSIGN \textit{collection}.
The updated query plan is:
	
\begin{lstlisting}
DISTRIBUTE-RESULT( $$13 )
UNNEST( $$13:child(treat(child(treat($$4, element_node), "bookstore"), element_node), "book") )
DATASCAN( collection("/books"), $$4 )
EMPTY-TUPLE-SOURCE
\end{lstlisting}

The finer grained tuples reduce the buffer size between operators during the query execution.
Note that the above rewrite rule allows VXQuery to process any amount of XML data provided that the largest XML document in the collection can fit in Hyracks' frame size.
This constraint can be further reduced to the largest subtree under the query path expression.
This is possible when the UNNEST \textit{child} expression is the consumer of a DATASCAN operator.
The \textit{child} expression can be merged into the DATASCAN operator to provide even smaller tuples.
The query plan is updated to show the DATASCAN operator has a third argument specifying the child path expression; 
the updated DATASCAN operator includes the path expression within the collection:
	
\begin{lstlisting}
DISTRIBUTE-RESULT( $$4 )
DATASCAN(collection("/books"),$$4,"/bookstore/book")
EMPTY-TUPLE-SOURCE
\end{lstlisting}

In addition to the improved pipeline, the DATASCAN operator offers a way to introduce partitioned parallelism simply by specifying VXQuery's partition details to this operator.
In VXQuery, data is partitioned among the cluster nodes.
Each node has a unique set of XML documents stored under the same directory specified in the \textit{collection} expression.
The Algebricks' physical plan optimizer uses these partitioned data properties details to distribute the query execution among the nodes.
For example, path "/books" defined in the \textit{collection} expression is located on each node and represents a unique set of XML documents for the query.
These partition properties are added to the DATASCAN operator although this is not shown in the query plan.
Adding these properties allows VXQuery to achieve partitioned parallel execution without any parallel programming.

\subsubsection{Replace Scalar with Aggregate Expressions}


The XQuery aggregate expressions (\textit{avg}, \textit{count}, \textit{max}, \textit{min}, and \textit{sum}) use a default scalar implementation in a normalized query plan.
This implies that the whole result is first stored in a sequence which is then processed to produce the aggregate.
Instead of creating this result sequence, we can match the XQuery aggregate expression with an Algebricks AGGREGATE operator.
When the Algebricks AGGREGATE operator is used with an XQuery aggregate expression, the result will be incremental aggregation instead of pre-collecting all records in the operator's buffer.
Consider the following query that counts the number of book elements in an XML collection and the query plan produced using the previous rules:
	
\begin{lstlisting}
count( 
  for $x in collection("/books")/bookstore/book 
  return $x
)
\end{lstlisting}

\begin{lstlisting}
DISTRIBUTE-RESULT( $$17 )
UNNEST( $$17:iterate($$16) )
ASSIGN( $$16:count(treat($$15, any_type)) )
SUBPLAN {
  AGGREGATE( $$15:create_sequence($$4) )
  DATASCAN(collection("/books"),$$4,"/bookstore/book")
  NESTED-TUPLE-SOURCE
}
EMPTY-TUPLE-SOURCE
\end{lstlisting}

The XQuery aggregate expression \textit{count} is within an ASSIGN \$\$16 operator. 
The SUBPLAN finds the bookstore nodes and uses an AGGREGATE \$\$15 operator to store them in a sequence.
However, there is no UNNEST directly above the SUBPLAN (as in Section \ref{SBR}) and thus the SUBPLAN cannot be removed. 
However, the scalar \textit{count} expression applies its calculation on the produced XQuery sequence to create \$\$16's value.
Instead, the aggregate \textit{count} expression can replace the \textit{create\_sequence} within the Algebricks AGGREGATE operator, thus performing aggregation incrementally instead of first generating a large XQuery sequence (in the buffer). 
The updated query plan becomes:

\begin{lstlisting}
DISTRIBUTE-RESULT( $$17 )
UNNEST( $$17:iterate($$16) )
SUBPLAN {
  AGGREGATE( $$16:count(treat($$4, any_type)) )
  DATASCAN(collection("/books"),$$4,"/bookstore/book")
  NESTED-TUPLE-SOURCE
}
EMPTY-TUPLE-SOURCE
\end{lstlisting}

The new plan keeps the pipeline granularity and enables partitioned aggregation processing.
An additional VXQuery rule annotates the AGGREGATE operator with local and global aggregate expressions, enabling the use of Algebricks' support for two-step aggregation -- each partition calculates its local aggregate result on its data and then transmits the result to a central partition for the global computation.
Thus, the partitioning also reduces the communication to improve the parallel processing.

\pagebreak
\subsubsection{Introduce the JOIN Operator}

In XQuery, two distinct datasets can be connected (matched) through a nested \textit{for} loop. 
The normalized query plan follows the same nested loop, which can be very expensive; we could do better by using a relational-style join. 
We note that Algebricks provides a JOIN operator as well as a set of language independent rewrite rules to optimize generic query plans. 
We can thus use these provided rewrite rules to translate the nested loop plan in to a join plan.
Consider a query that takes two bookstores (Ann and Joe) and finds books with the same title, and its query plan (note that the collections stay in dataflow order): 

\begin{lstlisting}
for $r in collection("/ann-books")/bookstore/book
for $s in collection("/joe-books")/bookstore/book
where $r/title eq $s/title
return $r
\end{lstlisting}

\begin{lstlisting}
DISTRIBUTE-RESULT( $$32 )
UNNEST( $$32:iterate($$31) )
SELECT( boolean(value-eq($$27, $$28)) )
ASSIGN( $$28:data(child($$26, "title")) )
ASSIGN( $$27:data(child($$13, "title")) )
DATASCAN(collection("/joe-books"),$$26,"/bookstore/book")
DATASCAN(collection("/ann-books"),$$13,"/bookstore/book")
EMPTY-TUPLE-SOURCE
\end{lstlisting}

In this example each dataset is identified and accessed by a DATASCAN operator, while the SELECT operator contains the condition for connecting the two datasets (which effectively will become the join condition).
The two ASSIGN operators (\$\$28 and \$\$27) find the title child node and return the string representation of the node. 
Three Algebricks language-independent rules are used to introduce the JOIN operator.
The first rule converts the nested DATASCAN operator into a cross product; it identifies that each data source is independent and adds the JOIN operator with a condition of true (basically a cross-product). 
The second Algebricks rule manipulates the DAG to push down operators that only affect one side of the join branch (selection, assign etc). 
The third rule then merges the SELECT and JOIN operators so the join condition (from the SELECT) is within the JOIN operator.
In the final plan, the JOIN operator has one branch from each data source, which allows each branch to be processed locally and then joined together globally.

\begin{lstlisting}[linewidth=3.43in]
DISTRIBUTE-RESULT( $$32 )
UNNEST( $$32:iterate($$31) )
JOIN( boolean(value-eq($$27, $$28)) )
{
  ASSIGN( $$28:data(child($$26, "title")) )
  DATASCAN(collection("/joe-books"),$$26,"/bookstore/book")
  EMPTY-TUPLE-SOURCE
} {
  ASSIGN( $$27:data(child($$13, "title")) )
  DATASCAN(collection("/ann-books"),$$13,"/bookstore/book")
  EMPTY-TUPLE-SOURCE
}
\end{lstlisting}

Going from here to the final logical plan does not require any custom VXQuery rules, but the physical plan needs more information to choose the most efficient join algorithm.
The equality comparison in our sample query allows the use of a more efficient partition-based algorithm.
If Algebricks understands the condition characteristics, it can chose an optimal hash-based join. 
For Algebricks to identify the join condition, this condition must be presented by a boolean Algebricks expression, in this case the Algebricks' \textit{equal} expression for a hash-based join. 
(Other Algebricks generic expressions include: \textit{and, or, not, less than, greater than, less than or equal, greater than or equal, not equal}.)
As the extraction of the XQuery's Effective Boolean Value of the value-comparison in the previous plan (\texttt{boolean(value-eq(...))}) is equivalent to Algebricks' \textit{equal} expression, we convert one to the other -- thus enabling Algebricks to identify the join.
After running the physical optimization rules the Algebricks expression is converted back to the original XQuery expressions for runtime execution.
As a result, Hyracks will use a Hybrid-Hash Join algorithm to achieve efficient partitioned parallelism.


\section{VXQuery Performance}\label{sec:performance}

To examine the scalability of our XQuery implementation we have performed an experimental evaluation using publicly available weather XML data.
We have also performed a comparison of VXQuery against two open source XML processors: Saxon \cite{kay2004saxon} and Apache MRQL \cite{mrql2014, fegaras2011xml}.

\subsection{Weather Data}

The NOAA website \cite{NOAA-GHCND} offers weather data via an XML-based web service. 
For our queries, we chose the Global Historical Climatology Network (GHCN)-Daily dataset that includes daily summaries of climate recordings. 
The core data fields report high and low temperatures, snowfall, snow depth, and rainfall. 
The complete data definition and field list can be found on NOAA's site \cite{NOAA-GHCND}. 
The date, data type, station id, value, and various attributes (i.e., measurement, source, and quality flags) are included for each weather report. 
In addition, a separate web service provides additional station data: name, latitude, longitude, and date of first and last reading.
The datasets used had four different sizes, ranging from 500MB up to 500GB.

\subsection{Queries}

We considered three basic types of XQuery  queries: selection, aggregation and join.

\textit{Selection:}
Query \ref{q:q00} shows historical data for the Key West International Airport, FL (USW00012836) station by selecting the weather readings on December 25th over the last 10 years.
Query \ref{q:q01} finds all readings that report an extreme wind warning. 
Such warnings occur when the wind speed exceeds 110 mph. 
(The wind measurement unit, tenths of a meter per second, has been converted to miles per hour.)

\begin{query}[Station-Date Weather History]{q:q00}
for $r in collection("/sensors")/dataCollection/data
let $datetime := dateTime(data($r/date))
where $r/station eq "GHCND:USW00012836" 
  and year-from-dateTime($datetime) ge 2003
  and month-from-dateTime($datetime) eq 12 
  and day-from-dateTime($datetime) eq 25
return $r
\end{query}

\begin{query}[Extreme Wind Warming]{q:q01}
for $r in collection("/sensors")/dataCollection/data
where $r/dataType eq "AWND" 
  and decimal(data($r/value)) gt 491.744
return $r
\end{query}

\textit{Aggregation:}
Query \ref{q:q02} finds the annual precipitation for Syracuse, NY using the airport weather station (USW00014771) for 1999. 
The precipitation is reported in tenths of an inch. 
Query \ref{q:q03} finds the highest recorded temperature in the weather data set. 
The Celsius temperature is reported in tenths of a degree.

\begin{query}[Syracuse Hancock Airport Annual Rainfall]{q:q02}
sum(
  for $r in collection("/sensors")/dataCollection/data
  where $r/station eq "GHCND:USW00014771" 
    and $r/dataType eq "PRCP" 
    and year-from-dateTime(dateTime(data($r/date))) eq 1999
  return $r/value
) div 10
\end{query}

\begin{query}[Highest Recorded Temperature]{q:q03}
max(
  for $r in collection("/sensors")/dataCollection/data
  where $r/dataType eq "TMAX"
  return $r/value
) div 10
\end{query}

\textit{Join:}
Query \ref{q:q04} finds all the weather readings for Washington state on a specific day in history: 1976/7/4. 
The date comparison uses the dateTime function to match the weather data provided.
Query \ref{q:q06} finds the highest recorded temperature (TMAX) for each station for each day during the year 2000.

\begin{query}[Specific Day-Station Readings]{q:q04}
for $s in collection("/stations")/stationCollection/station
for $r in collection("/sensors")/dataCollection/data
where $s/id eq $r/station 
  and (some $x in $s/locationLabels satisfies (
      $x/type eq "ST" and
      upper-case(data($x/displayName)) eq "WASHINGTON"))
  and dateTime(data($r/date))
      eq dateTime("1976-07-04T00:00:00.000")
return $r
\end{query}

\begin{query}[High Temperature per Station]{q:q06}
for $s in collection("/stations")/stationCollection/station
for $r in collection("/sensors")/dataCollection/data
where $s/id eq $r/station
  and $r/dataType eq "TMAX" 
  and year-from-dateTime(dateTime(data($r/date))) eq 2000
return ($s/displayName, $r/date, $r/value)
\end{query}

\textit{Join and Aggregation:}
Query \ref{q:q05} finds the lowest recorded temperature (TMIN) in the United States for 2001. 
In Query \ref{q:q07} we join two large relations, one that maintains the daily minimum temperature per station and one that contains the daily maximum temperature per station.
The join is on the station id and date and finds the daily temperature difference per station and returns the average difference over all stations.

\begin{query}[Min Temperature in the US]{q:q05}
min(
  for $s in collection("/stations")/stationCollection/station
  for $r in collection("/sensors")/dataCollection/data
    where $s/id eq $r/station
    and (some $x in $s/locationLabels satisfies 
        ($x/type eq "CNTRY" and $x/id eq "FIPS:US"))
    and $r/dataType eq "TMIN" 
    and year-from-dateTime(dateTime(data($r/date))) eq 2001
  return $r/value
) div 10
\end{query}

\begin{query}[Average Daily Temperature Differential]{q:q07}
avg(
  for $r_min in collection("/sensors_min")/dataCollection/data
  for $r_max in collection("/sensors_max")/dataCollection/data  
  where $r_min/station eq $r_max/station
    and $r_min/date eq $r_max/date
    and $r_min/dataType eq "TMIN"
    and $r_max/dataType eq "TMAX"
  return $r_max/value - $r_min/value
) div 10
\end{query}

\subsection{Experimental Results}

Our performance study explores VXQuery's ability to scale locally with the number of cores and then in a cluster with the number of nodes.
In the single node tests, the number of data partitions is varied to demonstrate nodes scaling up to the number of available cores.
For these tests, partitions represent data splits and each partition has a separate query execution thread.
In the cluster tests, the number of partitions per node has been fixed (to one partition per core) and only the number of nodes is varied.
The tests were all executed on an eight-node gigabit-connected cluster.
Each node has two dual-core AMD Opteron(tm) processors, 8GB of memory, and two 1TB hard drives.

\subsubsection{Single Node Experiments}

Our single node experiments used one cluster node and repeated each query five times.
The reported query time is an average of the last three runs.
(In our setting, the first two query executions times varied while the operating system cached the files.)
The first single node experiment compares VXQuery with
Saxon \cite{kay2008ten}, which is a highly efficient open source XQuery processor. 
The freely available Saxon Home Edition (SaxonHE 9.5) is typically limited to a single thread processing data that can fit into one fifth the size of the machine's memory. 
A group of weather stations were selected to create query results that fit these Saxon data restrictions.
The 584MB data set has been partitioned on a single hard drive.
The speed-up test keeps the total data set size constant while varying its number of data partitions and corresponding threads used to process the queries.

\begin{performancefigures}
{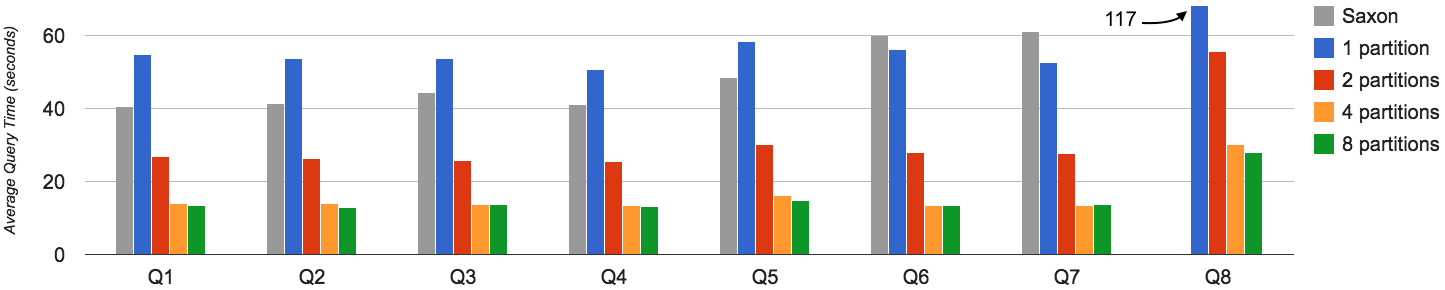}
\caption{Single node speed up comparison for Saxon and VXQuery (584MB dataset; varying VXQuery partitions).}
\label{fig:local_speed_up_584m}
\end{performancefigures}

Figure \ref{fig:local_speed_up_584m} shows the single node speed-up performance results for VXQuery and Saxon.
A single result per query is shown for Saxon since multi-threading is not available in the SaxonHE 9.5 version.
VXQuery outperforms Saxon when it uses two or more partitions.
The join queries (Query \ref{q:q04} through \ref{q:q07}) are translated into hash-based joins for VXQuery, thus giving better performance than Saxon's nested loop join.
Saxon's result for Query \ref{q:q07} is not reported since the large number of joined records caused it to never complete (and based on our other results, this query could take several months to complete).
For the rest of the queries (Query 1 through 7), when using 4 or more partitions, VXQuery performed on average about 3 times faster than Saxon.  
The VXQuery performance for 8 partitions is similar to the performance for 4 partitions, which is when the CPU becomes saturated. 

\begin{performancefigures}
{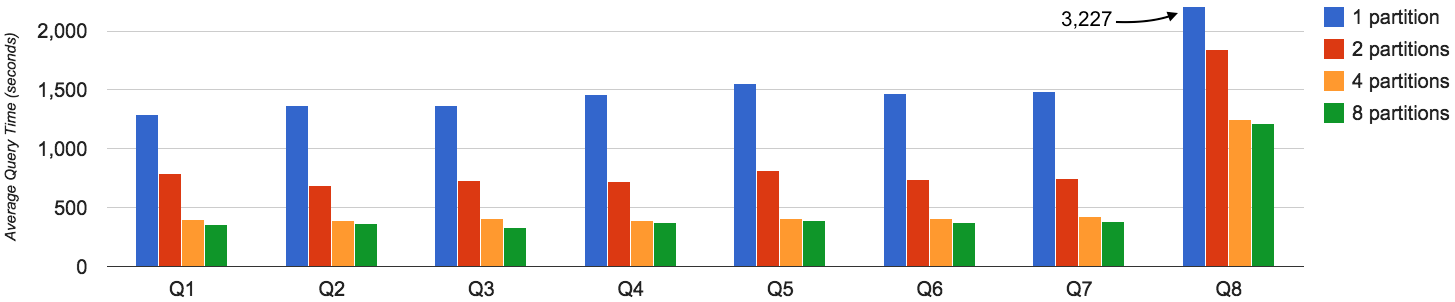}
\caption{Single node VXQuery speed up (15.2GB dataset).}
\label{fig:local_speed_up_16g}
\end{performancefigures}

To further test single node speed-up for VXQuery, we also used a dataset larger than the node's memory (8GB). 
For this we used a XML weather data subset, the GCOS Surface Network (GSN) stations containing 15.2GB of XML data. 
The results appear in Figure \ref{fig:local_speed_up_16g}.
As with the previous figure, VXQuery scales well up to the node's number of cores (4).
Similar to the single node 584MB experiments, the CPU is saturated when using 4 or more partitions. 
While profiling our experiments, we observed that VXQuery is CPU bound here, despite the larger data size, due to the overhead of parsing the XML document for each query. 
This is also evident from the improvement in performance when increasing threads.

\pagebreak
\subsubsection{Cluster Experiments}

Based on the single node speed-up results, the cluster experiments used eight nodes and four partitions per node.
The first cluster tests used the U.S. Historical Climatology Network (HCN) stations dataset which holds 57GB of XML weather data.
This dataset exceeds the available cluster memory when using less than eight nodes.
For each experiment, the dataset was equally divided among the nodes participating in the experiment.

\begin{performancefigures}
{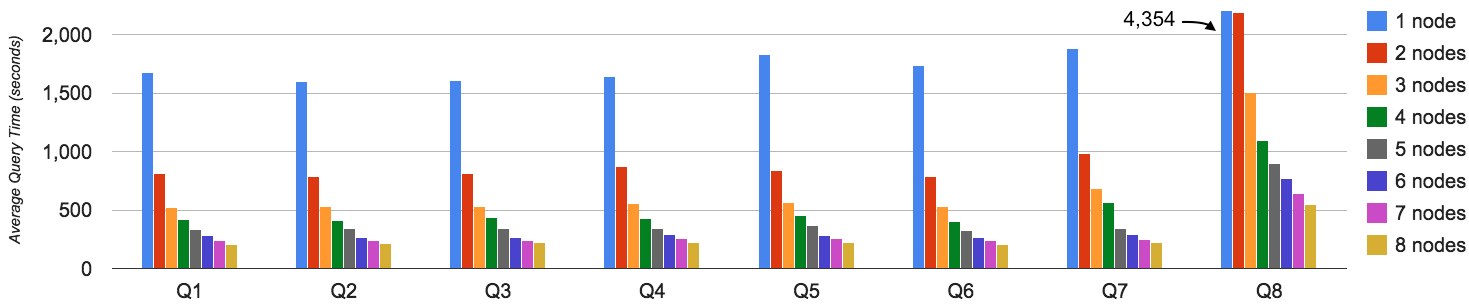}
\caption{VXQuery cluster speed up (57GB dataset).}
\label{fig:cluster_speed_up_60g_4p}
\end{performancefigures}

\begin{performancefigures}
{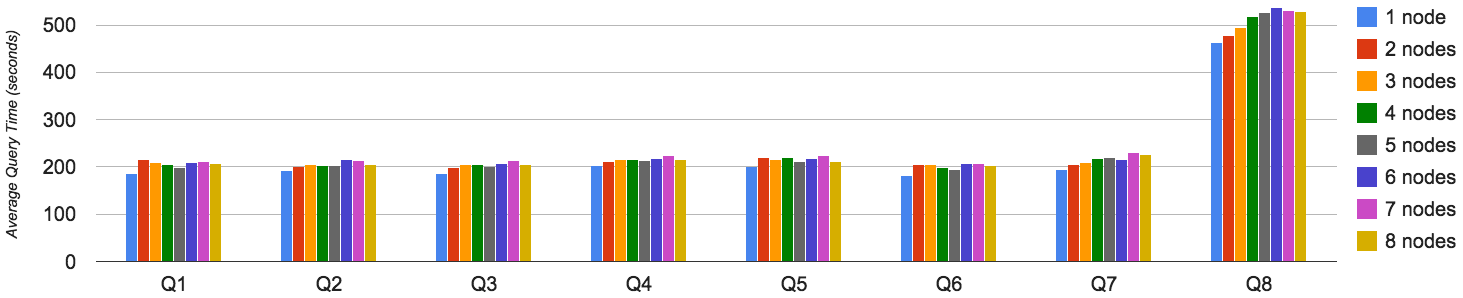}
\caption{VXQuery cluster scale up (7.2GB per node).}
\label{fig:cluster_scale_up_60g_4p}
\end{performancefigures}

The cluster speed-up results for VXQuery appear in Figure \ref {fig:cluster_speed_up_60g_4p}.
As can be observed, adding nodes to the cluster proportionally lowers the query time.
We next tested the scale-up characteristics of VXQuery. We started by using a dataset that fits in the memory of each node (i.e., 7.2GB of data per node). 
The results appear in Figure \ref{fig:cluster_scale_up_60g_4p}. 
While nodes are added to the query, the query time remains comparable, that is, the additional data workload is processed in the same amount of time.
VXQuery thus scales up well for Queries 1 through 7 on XML data.

The next scale-up test utilizes all 528GB of weather data; here each node has 66GB of data split evenly on two local disks. 
The results appear in Figure \ref{fig:cluster_scale_up_500g_4p}; VXQuery clearly scales-up well even for very large XML datasets.

\begin{performancefigures}
{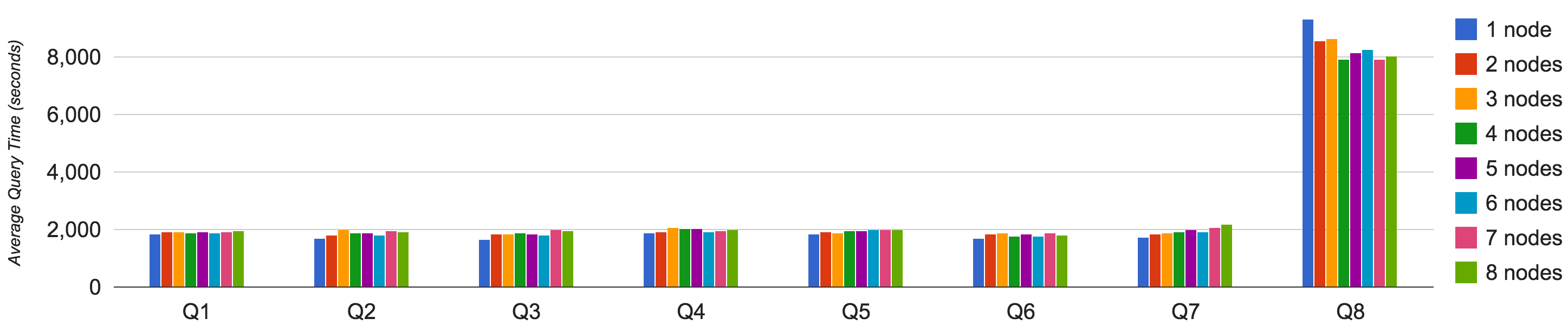}
\caption{VXQuery cluster scale up  (66GB per node).}
\label{fig:cluster_scale_up_500g_4p}
\end{performancefigures}

Our final experiment sought to evaluate VXQuery's performance against other open source parallel XQuery processors. 
Among them we chose Apache MRQL \cite{fegaras2011xml} as it was readily available. 
Using the HCN (57GB) dataset, we ran speed-up and scale-up tests for the same queries on Apache MRQL running on top of Hadoop 1.2.1 using MapReduce.
Hadoop was configured with a 128MB block size and a replication factor of 1 (to reduce space on the cluster).
VXQuery outperforms Apache MRQL on all queries through both scale-up and speed-up.
Figures \ref{fig:mrql_speed_up_60g} and \ref{fig:mrql_scale_up_60g} show the Apache MRQL query times; for comparison the respective VXQuery query times (from Figures \ref {fig:cluster_speed_up_60g_4p} and \ref{fig:cluster_scale_up_60g_4p}) are also shown (via the circles part way up each bar).
VXQuery's performance advantage comes in part from reading and parsing XML about two times faster than Apache MRQL.
In addition, its richer set of operators allows for better performance. 
For example, XQuery utilizes a Hybrid Hash Join algorithm that keeps a partition in main memory. 
Instead, by using MapReduce, Apache MRQL divides the join responsibility: the partitioning is done by the mapper, while the reducer joins the individual partitions. 
These two steps do not share state, yielding a traditional Grace Hash Join.
On average over all experiments, VXQuery is two and half times faster than Apache MRQL on Hadoop, validating the fact that building XQuery on top of a dataflow environment like Hyracks provides more opportunities for optimization and parallelism.

\begin{performancefigures}
{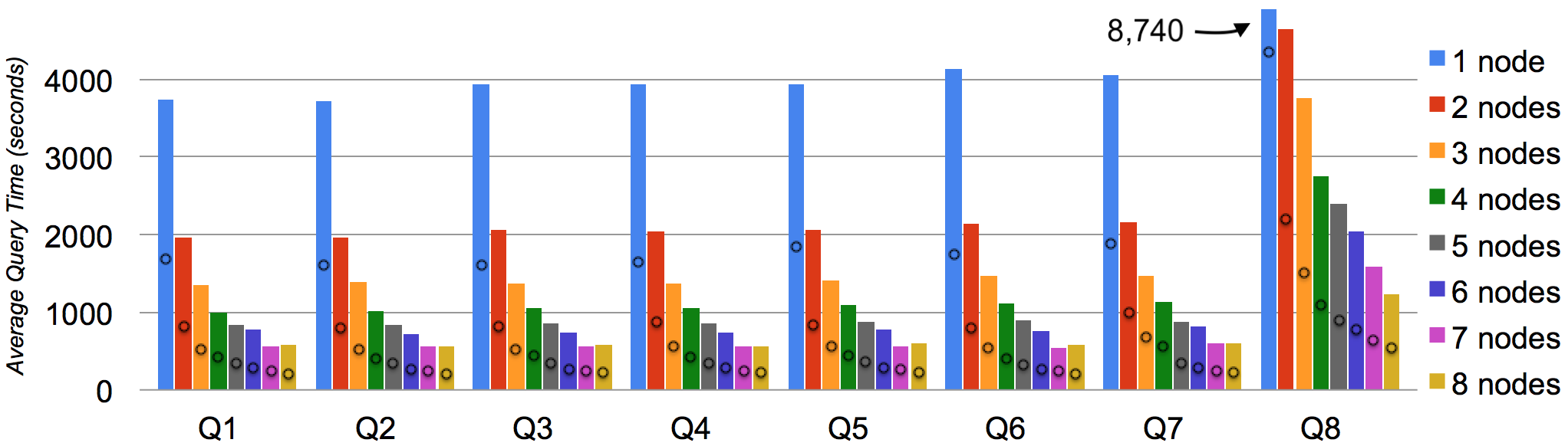}
\caption{MRQL cluster speed up (57GB dataset); circles mark the respective VXQuery times.}
\label{fig:mrql_speed_up_60g}
\end{performancefigures}

\begin{performancefigures}
{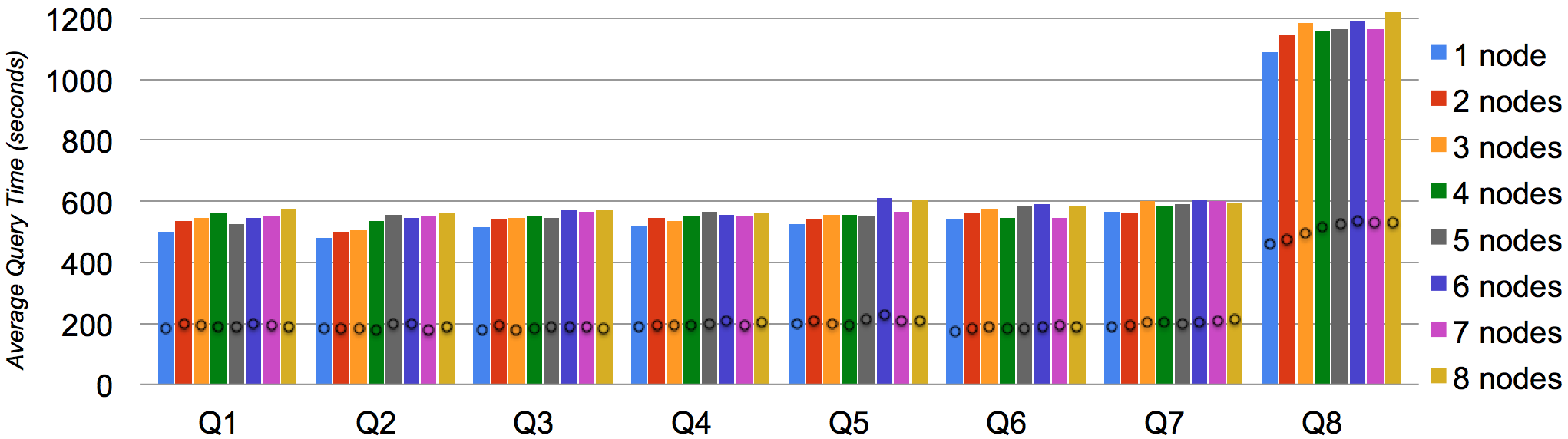}
\caption{MRQL cluster scale-up (7.2GB per node); circles mark the respective VXQuery times.}
\label{fig:mrql_scale_up_60g}
\end{performancefigures}

\section{Conclusions}\label{sec:conclusion}

VXQuery is a scalable open-source XQuery processor that we have built on top of Hyracks and Algebricks. 
We have described its implementation, including the XML 'data model'-dependent rewrite rules. These rules facilitate existing, 'data model'-independent Algebricks optimizations that serve to create efficient and parallel Hyracks jobs.
We have demonstrated using a real 500GB dataset that VXQuery can scale out to the number of nodes available on a cluster for various XML selection, aggregation, and join queries.
Comparatively, VXQuery is about 3x faster than Saxon on a single node and around 2.5x faster than Apache MRQL on a cluster in terms of scale-up and speed-up.
The source code is available as an Apache Software Foundation \cite{vxquery2014}; the current release contains approximately 100K LOC.
The planned next steps are to increase coverage of XQuery 1.0 features and to add XQuery 3.0 features that support the analysis of Big Data, such as the ^group by^ and ^window^ clauses.

{\bf Acknowledgments:}
The work was supported in part by Google Summer of Code (2012 and 2013) and by NSF grants IIS-0910859, IIS-0910989, CNS-1305253 and CNS-1305430. 

\bibliographystyle{abbrv}
\bibliography{sigproc}
\balance
\end{document}